\documentclass[prl,twocolumn,showpacs]{revtex4} %showpacs,preprint,showkeys

\usepackage{latexsym}
\usepackage{graphicx}
\begin{document}
\title{Spatially resolved study of backscattering in the quantum spin Hall state}

\date{\today}
%\author{Markus K\"{o}nig}
%\author{Matthias Baenninger}
%\affiliation{Department of Physics, Stanford University, Stanford, CA 94305, USA}
%\affiliation{Stanford Institute for Materials and Energy Sciences, SLAC National Accelerator Laboratory, 2575 Sand Hill Road, Menlo Park, CA 94025, USA}
%\author{Andrei~G.~F.~Garcia}
%\affiliation{Department of Physics, Stanford University, Stanford, CA 94305, USA}
%\author{Nahid Harjee}
%\affiliation{Department of Electrical Engineering, Stanford University, Stanford, CA 94305, USA}
%\author{Beth L. Pruitt}
%\affiliation{Department of Mechanical Engineering, Stanford University, Stanford, CA 94305, USA}
%\author{C. Ames}
%\author{Philipp Leubner}
%\author{Christoph Br\"{u}ne}
%\author{Hartmut Buhmann}
%\author{Laurens W. Molenkamp}
%\affiliation{Physikalisches Institut(EP 3), Universit\"{a}t W\"{u}rzburg, Am Hubland, 97074 W\"{u}rzburg, Germany}
%\author{David Goldhaber-Gordon}
%\email{goldhaber-gordon@stanford.edu}
%\affiliation{Department of Physics, Stanford University, Stanford, CA 94305, USA}
%\affiliation{Stanford Institute for Materials and Energy Sciences, SLAC National Accelerator Laboratory, 2575 Sand Hill Road, Menlo Park, CA 94025, USA}

\author{Markus K\"{o}nig$^{1,2}$, Matthias Baenninger$^{1,2}$, Andrei~G.~F.~Garcia$^{1}$, Nahid Harjee$^{3}$, Beth L. Pruitt$^{4}$, C. Ames$^{5}$, Philipp Leubner$^{5}$, Christoph Br\"{u}ne$^{5}$, Hartmut Buhmann$^{5}$, Laurens W. Molenkamp$^{5}$, David Goldhaber-Gordon$^{1,2}$}
\email{goldhaber-gordon@stanford.edu}
\affiliation{$^1$ Department of Physics, Stanford University, Stanford, CA 94305, USA\\$^2$ Stanford Institute for Materials and Energy Sciences, SLAC National Accelerator Laboratory, 2575 Sand Hill Road, Menlo Park, CA 94025, USA,\\$^3$ Department of Electrical Engineering, Stanford University, Stanford, CA 94305, USA\\$^4$ Department of Mechanical Engineering, Stanford University, Stanford, CA 94305, USA\\$^5$ Physikalisches Institut (EP3), Universit\"{a}t W\"{u}rzburg, Am Hubland, 97074 W\"{u}rzburg, Germany}

\begin{abstract}
The discovery of the Quantum Spin Hall state, and topological insulators in general, has sparked strong experimental efforts. Transport studies of the Quantum Spin Hall state confirmed the presence of edge states, showed ballistic edge transport in micron-sized samples and demonstrated the spin polarization of the helical edge states. While these experiments have confirmed the broad theoretical model, the properties of the QSH edge states have not yet been investigated on a local scale.

Using Scanning Gate Microscopy to perturb the QSH edge states on a sub-micron scale, we identify well-localized scattering sites which likely limit the expected non-dissipative transport in the helical edge channels. In the micron-sized regions between the scattering sites, the edge states appear to propagate unperturbed as expected for an ideal QSH system and are found to be robust against weak induced potential fluctuations.
\end{abstract}


%\keywords{}

\maketitle

Ever since the prediction and realization of topological insulators, this new class of material has not only been studied for fundamental scientific reasons, but also garnered significant interest due to potential applications. In particular, the edge channels of the Quantum Spin Hall (QSH) state are considered a promising candidate for low-power information processing because of the expected dissipationless transport in the one-dimensional helical edge channels. In the Quantum Spin Hall state~\cite{Kane05, Bernevig06}, backscattering between the counterpropagating channels in one helical edge state is expected to be suppressed as long as the Fermi level is located in the bulk gap, time-reversal symmetry is preserved, and interactions are weak~\cite{Wu06, Xu06}. In HgTe quantum well structures, the QSH state was predicted to exist if the thickness of the quantum well layer exceeds a critical value of $d_{QW} \approx 6.3$~nm~\cite{Bernevig06b}. Following the initial observation of the QSH state in HgTe~\cite{Koenig07}, the edge-state nature of transport~\cite{Roth09} and the spin-polarization of the edge states~\cite{Bruene12} were experimentally demonstrated in the same material. Thus it can serve as a model system for the investigation of the QSH state even though its properties make room-temperature applications appear unlikely. The good agreement between the experimental results and theoretical predictions based on the Landauer-B\"{u}ttiker formalism confirmed that transport in the helical edge channels is indeed ballistic over short distances. However, conductance values indicative of ballistic transport were only observed on edges shorter than a few microns~\cite{Koenig07, Koenig08, Roth09}.  In Ref.~\citenum{Roth09}, a measured nonlocal conductance value could be explained within the Landauer-B\"{u}ttiker formalism, assuming a single scattering site that fully equilibrates the counterpropagating channels along one particular segment of the device edge. The equilibration was attributed to decoherence of the helical states in a metallic region coupled to the edge states. Such a metallic puddle could form in an inhomogeneous potential landscape introduced by sample growth or processing. In multi-terminal devices, the metallic contacts to the QSH edge states can be treated as macroscopic regions causing equilibration, explaining the experimentally observed quantized resistance of the QSH state~\cite{Koenig07, Koenig08, Roth09}. Besides dephasing in metallic regions, various other ways to induce backscattering in a helical edge state have been studied theoretically, including mechanisms based on magnetic~\cite{Maciejko09} or non-magnetic~\cite{Schmidt12} impurities, spin-orbit interaction~\cite{Stroem10, Vaeyrynen11}, or phonons~\cite{Budich12}. Against this background of theoretical effort to investigate scattering mechanisms in the QSH state and the obvious presence of backscattering in transport measurements, more detailed experimental investigations of scattering are urgently needed.

\section*{Samples and Measurement Technique}

A promising approach to study backscattering in the helical edge states on a local scale is to generate small, local potential fluctuations and test whether they perturb transport in the edge channels. To achieve this goal, one must induce a local potential perturbation with precise control over its position and strength. This allows either modulation of the strength of a pre-existing scattering site or the generation of an artificial scattering site. In Scanning Gate Microscopy (SGM)~\cite{Eriksson96, Topinka00}, a charged tip is scanned above the device of interest and the effect of the induced potential fluctuation on the conductance in the device is monitored. Over the past few years, Scanning Gate Microscopy has been used to investigate transport phenomena in a variety of low-dimensional systems like quantum dots~\cite{Pioda04, Schnez10, Huefner11, Aoki12}, quantum point contacts~\cite{Schnez11} and two-dimensional electron gases~\cite{Topinka01, Jura07, Jura09, Jura10}. Of particular relevance to QSH edge state transport, SGM has been applied to one-dimensional systems like carbon nanotubes~\cite{Bachtold00, Woodside02} and nanowires~\cite{Bleszynski08} where it was used to identify and manipulate localized states controlling the transport through the device.

\begin{figure}
		\includegraphics[width=0.9\linewidth]{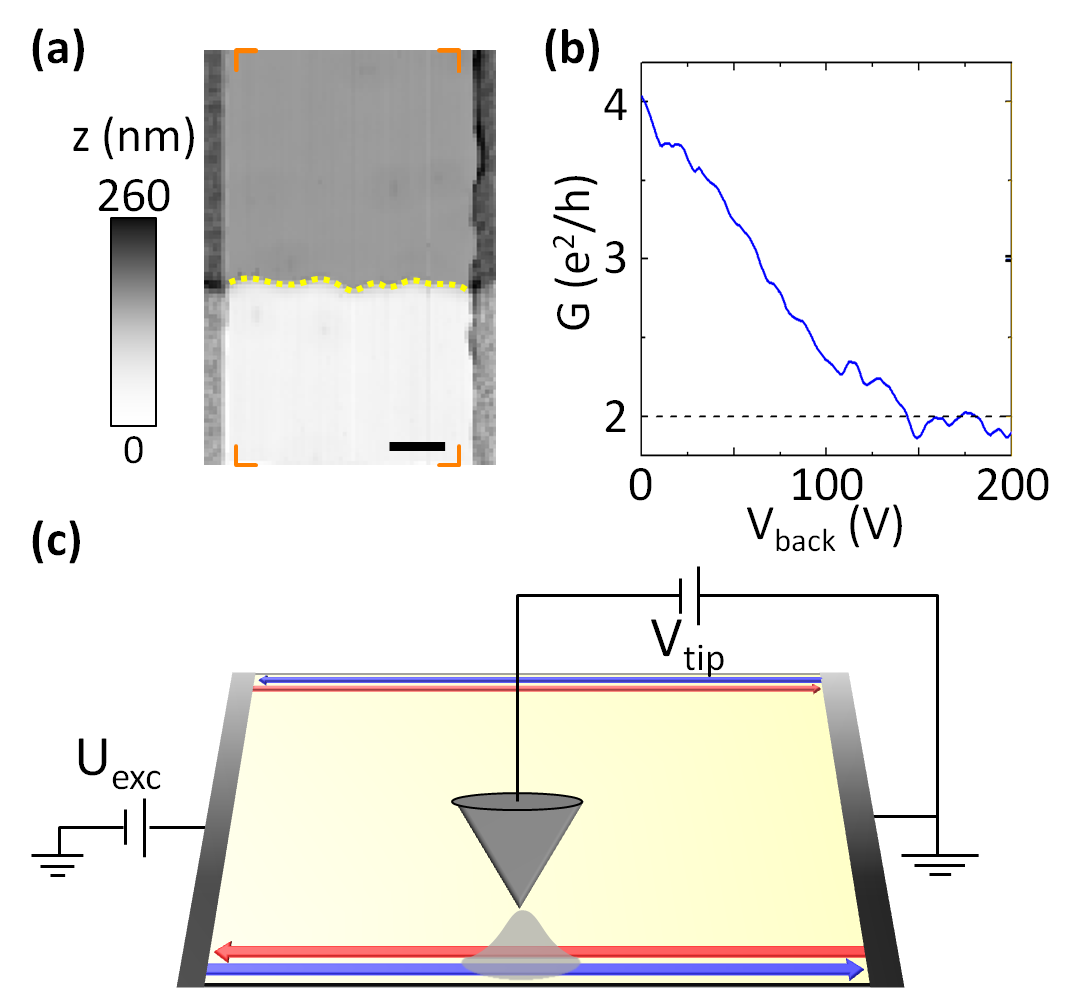}
		\caption{(Color) \textbf{(a)} In-situ AFM scans (scale bar $1~\mu$m) allow for precise alignment of the tip to the device. The mesa extends upward from the dotted yellow line. The grainy areas at the left and right edge, respectively, are the Ohmic contacts to the quantum well. The orange markers demarcate the scan window used for the conductance maps in Figs.~2 and 3. \textbf{(b)} A back-gate electrode is used to tune the device into the QSH regime with $G\leq 2e^2/h$. \textbf{(c)} Illustration of the experimental configuration for SGM studies of the QSH state.}
	%\label{Fig1}
\end{figure}

Our devices were fabricated from undoped HgTe/Hg$_{0.3}$Cd$_{0.7}$Te quantum well structures with a nominal HgTe layer thickness $d_{QW}=8$~nm. The devices were patterned using optical lithography, ion-milling to define mesas, and evaporation of In/Au ohmic contacts to the quantum well. Fig.~1a shows a section of a typical device in a topographic image taken by atomic force microscopy (AFM). Our home-built SGM probes feature self-sensing piezoresistive deflection readout~\cite{Harjee10} which allows for precise in-situ alignment of the tip to the device. The separation between the contacts in the transport direction is $5~\mu$m, and the lateral mesa width is $150~\mu$m. As this width is several orders of magnitude greater than the predicted extension of the QSH edge states into the bulk~\cite{Zhou08, Wada11}, any effect of inter-edge tunneling~\cite{Hou09, Stroem09, Schmidt11} can be ruled out. More relevant to our SGM experiments, the large width of the device ensures that the tip-induced potential perturbation only affects the transport along the edge located within the scan window, while the far edge remains unaffected and thus provides a constant contribution to the device conductance. We perform our experiments in a two-terminal configuration using standard lock-in technique ($f\approx 830$~Hz, $U_{exc}=100~\mu$V) at $T \approx 2.7$~K. Despite the absence of intentional doping in the heterostructure, the studied quantum well structures consistently show a finite density of $p$-type bulk carriers, typically on the order of a few times $10^{10}~\rm{cm}^{-2}$. For the device discussed extensively in this paper, we found a density $p \approx 2\times 10^{10}~\rm{cm}^{-2}$, prior to gating~\cite{Supp}. Fig.~1b shows the device conductance as a function of the voltage applied to the back gate electrode which is used to tune the Fermi level in the device. The observed saturation of the conductance slightly below $G=2e^2/h$ for $V_{back} > 150$~V suggests that the sample is in the QSH state and transport happens only along the edges. The basic concept of SGM measurements on a sample in the QSH state is illustrated in Fig.~1c: a DC voltage $V_{tip}$ is applied to the tip located slightly above the sample surface (in our experiments, height $h=90$~nm). The electric field from the tip produces a local modulation of the potential in the sample, below the tip. In our SGM experiments, we induce such a potential perturbation near the device edge with the goal of introducing and controlling backscattering of the helical edge channels. We studied several devices with similar properties, formed from different heterostructures, and all devices showed comparable results except as noted. In this paper, we focus our discussion on one particular device, which shows the same qualitative features as other devices as well as striking and enlightening behavior seen only on this device. Unfortunately, a tip crash damaged the device before we could collect a full data set for the sample in the QSH regime. However, earlier measurements on the same device in the presence of bulk carriers complement the data and allow for a detailed analysis. While the limited amount of data does not allow for a fully conclusive study of scattering mechanisms affcting the QSH edge channels, our results nonetheless provide a substantial step forward over the available transport experiments.

\section*{Scanning Gate Microscopy in the QSH regime}

\begin{figure}
		\includegraphics[width=\linewidth]{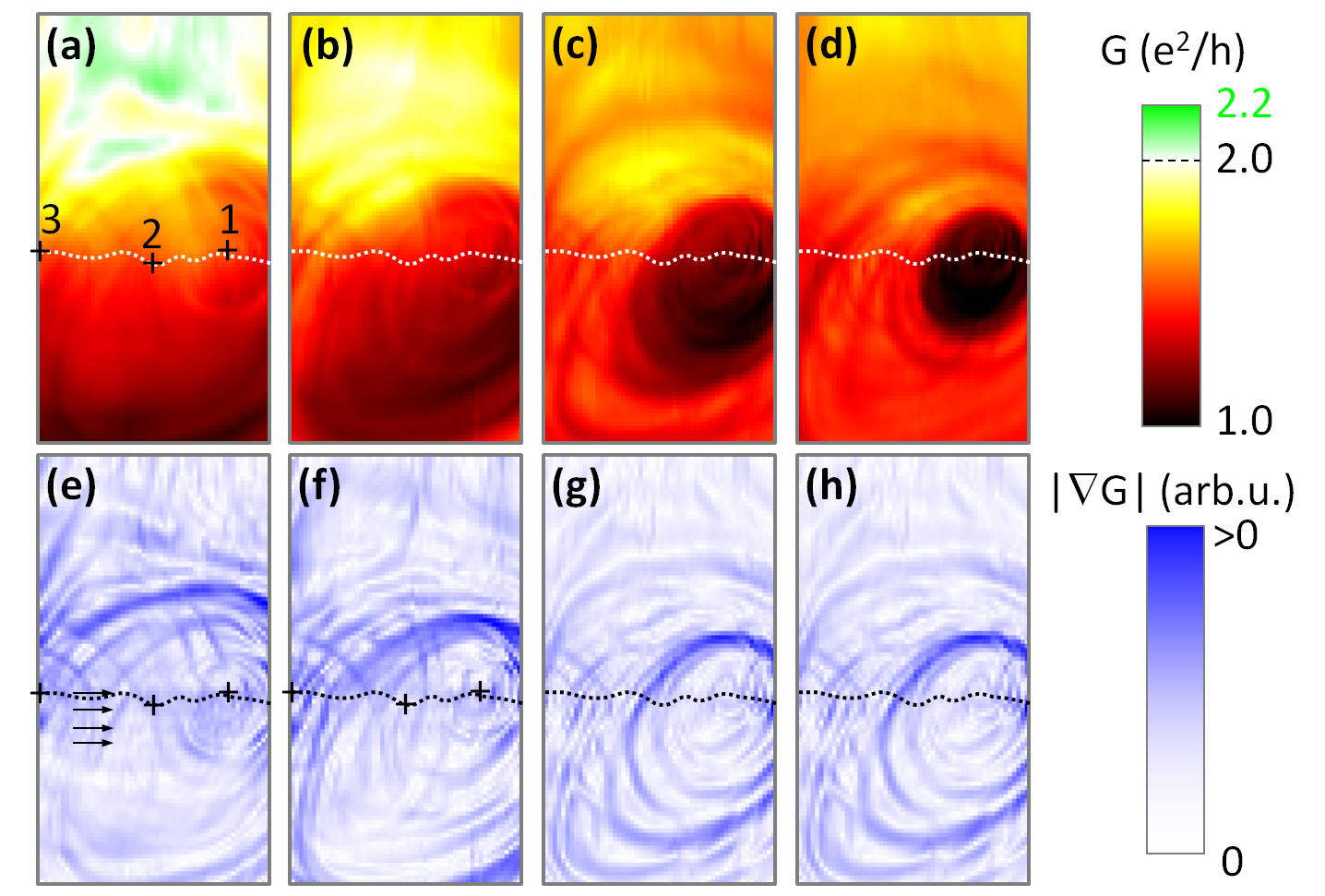}
		\caption{(Color) \textbf{(a)~-~(d),} Conductance maps $G(x,y)$ with $V_{back}=+200$~V for $V_{tip} = \rm{(\textbf{a})}-14.0~\rm{V}, \rm{(\textbf{b})}-12.5~\rm{V}, \rm{(\textbf{c})}-11.0~\rm{V}, \rm{(\textbf{d})}-9.5~\rm{V}$. The dotted lines indicate the edge of the device. The labeled crosses mark the positions of the scattering sites. \textbf{(e)~-~(h),} Gradient $|\nabla G(x,y)|$ of panels \textbf{(a)~-~(d)}.}
	%\label{Fig2}
\end{figure}

In the QSH regime at $V_{back} = +200$~V, we observe a strong modulation of the conductance in the SGM maps on this device (Fig. 2a-d): as a function of the tip position, the conductance varies approximately from $2e^2/h$ (similar to the conductance value when the unbiased tip is far away) down to $1e^2/h$. The rather smooth variation of $G$ over a length scale of several microns in tip position perpendicular to the edge of the device, most clearly visible in Fig.~2a, is mostly caused not by a local gating by the tip but rather by long-range gating by the large conductive area of the cantilever biased to the same voltage as the tip. In contrast, the superimposed modulations on length scales below 1 micron, highlighted by taking the gradient of conductance with tip position in Fig.~2e-h, are associated with local gating by the SGM tip, showing the effect of a well-localized potential perturbation on the edge state conductance. The suppression of conductance by $|\Delta G| \approx 1e^2/h$ by local gating suggests that transport at one of the two edges is fully suppressed by the tip potential. This result supports our assumption that the sample is in the QSH state, with transport occurring primarily along the edges of the device. For a strongly negative tip voltage $V_{tip}=-14.0$~V (Fig.~2a), regions with $G>2e^2/h$ can be observed in the conductance map. This cannot be explained solely by transport in the QSH edge states, and we attribute the excess conductance to the tip-induced emergence of bulk conductance (see Supplemental Material~\cite{Supp}). 

As noted above, some prior transport measurements~\cite{Roth09} may be understood by invoking a single fully edge-equilibrating metallic region along a single edge. This would decrease the conductance of that edge by a factor of 2, from $1e^2/h$ to $0.5e^2/h$. Based on a simple cartoon picture for our SGM experiments, we expected the tip potential to induce such a local metallic region below the tip, so we were surprised to see a stronger suppression than $0.5e^2/h$ associated with a single site. Several theoretical models have been proposed that could in principle lead to full suppression of edge conduction~\cite{Maciejko09, Stroem10, Qi08, Timm11, Ilan12}. However, as neither of those theoretical predictions seems sufficient to fully explain the observed behavior~\cite{Supp} and we found only a single occurrence of full suppression of edge state transport in all of the measured devices, we focus now on the more commonly seen features in the SGM scans.

Besides the pronounced conductance suppression attributed to a single site, the conductance maps show multiple sets of concentric rings representing a reduced conductance. Around site 1, the rings are superimposed on the strong conductance modulation (the full suppression). Two more sets of concentric rings are centered around sites 2 and 3, respectively, in Fig.~2. While these features are partly masked in the conductance maps by the dominant conductance modulation originating from site 1, they are clearly visible when the gradient of the conductance is plotted (Fig.~2e-h). For all three scattering sites, the conductance modulation associated with the sharp rings of conductance suppression is on order $0.1 e^2/h$. Circular patterns with a similar conductance modulation were found in all devices we studied.

In general, circular patterns in SGM maps are a signature that transport is a function of the potential at one sensitive site, the common center of the circles. This can be understood intuitively: when the tip moves along one of the circles, the tip-induced potential at the circle's center does not change and the conductance remains unaffected. However, when the tip moves toward or away from the sensitive site, the potential at that site and consequently the conductance change. The canonical example of this behavior is a quantum dot, where the local potential at the dot determines the occupancy, and the conductance peaks in Coulomb-blockaded transport show up as rings in SGM experiments~\cite{Woodside02, Bleszynski08, Schnez10}. The observation of ring-like patterns in our data suggests that the suppression of conductance in the QSH edge channel is linked to individual sites located at or near the physical edge of the device, though in place of Coulomb blockade we below suggest another mechanism dependent on the local potential. 
In our experiments, the rings of conductance suppression appear elongated due to spatial variation in the dielectric environment: when the tip is located above the mesa, the tip potential experienced at the scattering site is screened by the Hg$_{0.3}$Cd$_{0.7}$Te cap layer in the heterostructure with $\epsilon_{\rm{HgCdTe}} = 12.7$ whereas for a tip position not above the mesa the relevant dielectric constant should be between $\epsilon_{\rm{HgCdTe}}$ and $\epsilon_{\rm{vacuum}} = 1$.

The magnitude and other experimental signatures of conductance modulation resulting in multiple rings about the three indicated scattering sites may be understood if these sites are associated with small bulk conducting regions adjoining the edge and serving as sources of decoherence~\cite{Roth09}. For simplicity, we will discuss transport only through the helical edge state we scan over and omit the contribution of $G = 1e^2/h$ by the far edge.
A macroscopic metallic contact well-coupled to the edge is expected to fully equilibrate the counterpropagating edge states as noted above. As there is no relationship between the spin orientation of an electron impinging on this contact and one emerging back into the edge, the electron should emerge with equal probability into the forward-moving (say, spin up) and backward-moving (say, spin down) channel, respectively. Consequently, conductance along the edge should be reduced from $G=1e^2/h$ to $G=0.5e^2/h$. A metallic region of intermediate size could also reduce the conductance if the edge states can lose their spin information during interaction with the charge carriers in this metallic puddle. In Ref.~\citenum{Roth09}, electron-electron interaction or electron-phonon interaction were given as possible sources for this ``decoherence" of the helical edge state, though the role of these particular types of interaction has not been discussed quantitatively in the model. The amount of conductance decrease depends on the exact size of the dephasing region and the strength of the dephasing process. For a finite-sized metallic region, resonances in dephasing should enhance the suppression of the edge state conductance for particular sizes of the dephasing region. The conductance suppression caused by this resonant process is predicted to be around $0.1e^2/h$~\cite{Roth09}, comparable to the size of the modulation observed in our SGM maps. In the experimental data accompanying the model put forward in Ref.~\citenum{Roth09}, however, similar resonances were not observed, pointing towards the presence of a metallic puddle resulting in strong decoherence of the QSH edge state, e.g., due to a very large size.

Our experimental results indicate that backscattering of the QSH edge state is caused by well-localized scattering sites along the edge of the device. The observed pattern in the conductance modulation is consistent with the theoretical model described above based on partial dephasing of the edge state in a small metallic region at each of these sites. Subsequently, we will use this picture to interpret the backscattering observed in our measurements. When either the tip position or the tip voltage is changed, the potential at the puddle will vary, modulating the puddle size and possibly the strength of the dephasing mechanism. This should tune the dephasing process in the metallic region in and out of resonance, resulting in an oscillatory conductance modulation. 
Electrostatic simulations show that the size of a metallic region - either induced by the tip potential itself or already existing in the sample - with a charge density exceeding a certain threshold value can be tuned by a few hundred~nm within the tip voltage range used in our experiments~\cite{Supp}. At the same time, the maximum density in the puddle, which should affect the strength of the dephasing mechanism, changes with the tip voltage as well. This variation in the size and strength of the dephasing region is expected to result in multiple resonances~\cite{Roth09} as seen in our experimental data.

Our explanation for the observed backscattering is corroborated by additional features which appear at strongly negative tip voltages $V_{tip}\leq-12.5$~V. The maps of $|\nabla G|$ show multiple lines running approximately parallel to the mesa edge along the entire length of the device (indicated by arrows in Fig.~2e). This pattern of the gradient corresponds to an oscillatory conductance modulation as a function of the tip position perpendicular to the mesa edge. We can rule out simple instrumental artifacts as follows: The SGM data are recorded during line scans in the direction of the observed conductance modulation, so a slow drift in the sample conductance can be ruled out as an explanation. Instrumental oscillations on a timescale comparable to the measurement time between individual lines cannot account for the observed pattern either as the lines only appear near the edge, evolve consistently with the applied tip voltage~\cite{Supp}, and only occur for particularly large tip voltages. Though the amplitude of this conductance modulation is less than $0.05 e^2/h$ - weaker than the effect caused by the localized scattering sites - it can be explained by the same mechanism. For a sufficiently strong potential perturbation induced by the SGM tip, a metallic puddle (here $p$-type because of the negative tip voltages) is induced at an otherwise unperturbed section of the edge. For a dephasing region of appropriate size, resonant suppression of conductance should be seen. These lines are only visible for much stronger tip voltages $V_{tip}\leq-12.5$~V compared to the ring features described above. This difference in tip voltage required to cause backscattering can provide a measure for the strength of the intrinsic potential fluctuations and the robustness of the otherwise unperturbed edge states, as discussed in more detail below. While the SGM maps for the device in the QSH regime provide clear evidence for the presence of scattering sites, experimental data for this transport regime are only available for a limited tip voltage range, so we move next to conditions of mixed bulk and edge conduction. 

\section*{Coexistence of edge and bulk conductance}

\begin{figure*}
		\includegraphics[width=0.9\linewidth]{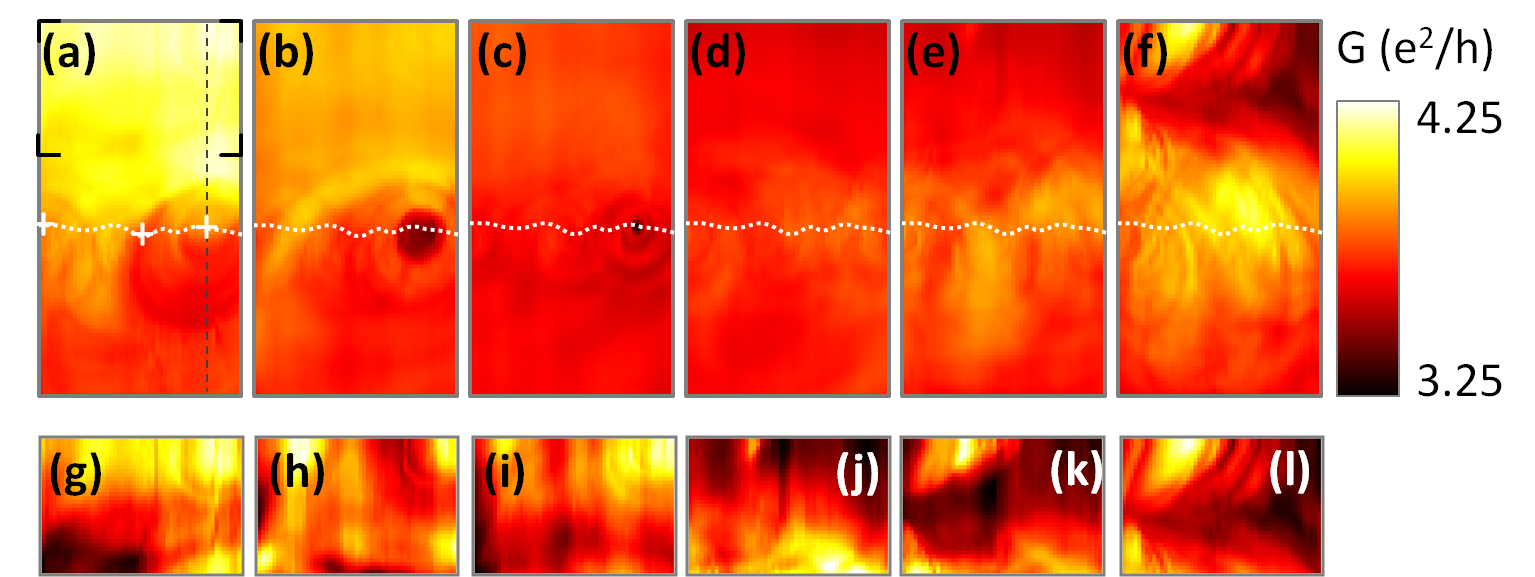}
		\caption{\label{fig:wide}(Color) \textbf{(a)~-~(f)} Conductance maps $G(x,y)$ taken at $V_{back}=0$  with tip voltages $V_{tip} =$ (\textbf{a}) -10.0~V, (\textbf{b}) -6.5~V, (\textbf{c}) -3.5~V, (\textbf{d}) +3.5~V, (\textbf{e}) +6.5~V and (\textbf{f}) +10.0~V, respectively. The vertical line in (\textbf{a}) indicates the position of the line scans shown in Fig.~4. \textbf{(g)~-~(l)}: Transport signatures of localized states in the bulk (window indicated in (a), color scale ($\Delta G$) is saturated for each scan individually): (\textbf{g}) $V_{tip}=-10.0\rm{V}, \Delta G=0.25e^2/h$; (\textbf{h}) $V_{tip}=-8.0\rm{V}, \Delta G=0.15e^2/h$; (\textbf{i}) $V_{tip}=-6.5\rm{V}, \Delta G=0.15e^2/h$; (\textbf{j}) $V_{tip}=+6.5\rm{V}, \Delta G=0.20e^2/h$; (\textbf{k}) $V_{tip}=+8.0\rm{V}, \Delta G=0.40e^2/h$; (\textbf{l}) $V_{tip}=+10.0\rm{V}, \Delta G=0.85e^2/h$.}
	%\label{Fig3}
\end{figure*}

At $V_{back}=0$, the sample displays a conductance $G\approx 4 e^2/h$. The conductance value larger than $2 e^2/h$ indicates that the device must have some bulk conductance. However, we expect QSH edge states to coexist with bulk carriers and contribute considerably to the total conductance. In this regime, the strongest local gating effect close to the edge reduces the conductance by $0.5 e^2/h$ (Fig.~4a). As this conductance modulation can be induced within a few hundred~nm of the edge, it is likely associated with an edge state and constitutes a lower limit for the conductance of a $5~\mu$m stretch of a single edge state under these slightly $p$-type conditions. This would be consistent with a reduced edge conductance in comparison to the QSH regime, possibly because coexisting bulk carriers allow backscattering of the edge states in the regions between the local scattering sites identified earlier. The reduced edge state conductance compared to the QSH regime at $V_{back}=+200$~V further confirms that the conductance along the edge is indeed due to QSH edge states and not caused by trivial edge currents as they may occur due to an inhomogeneous potential landscape. At $V_{tip}=-3.5$~V, the dominant conductance modulation $|\Delta G|\approx 0.5e^2/h$ occurs when the tip is located directly above the scattering site - within 125~nm (half width at half maximum), giving an estimate for the size of the dephasing region (Fig.~4a). The identification of this feature as suppression of edge state conductance is supported by the observation of an associated ring pattern centered at the edge of the mesa (Fig.~3a-c), similar to those observed in the QSH regime. We can see two more sets of rings, and all three sets are centered around the same three locations as in the QSH regime and presumably to the same scattering sites. Thus, we can use these data for a more complete analysis of the properties of the dephasing regions. While the suppression caused by site 1 is unusually strong and the underlying mechanism is not understood yet, the superimposed resonances are qualitatively comparable to the features centered at scattering sites 2 and 3. As the conductance modulation associated with site 1 is most clearly visible in the SGM maps, we use scattering site 1 for further analysis. 

\begin{figure}
		\includegraphics[width=0.9\linewidth]{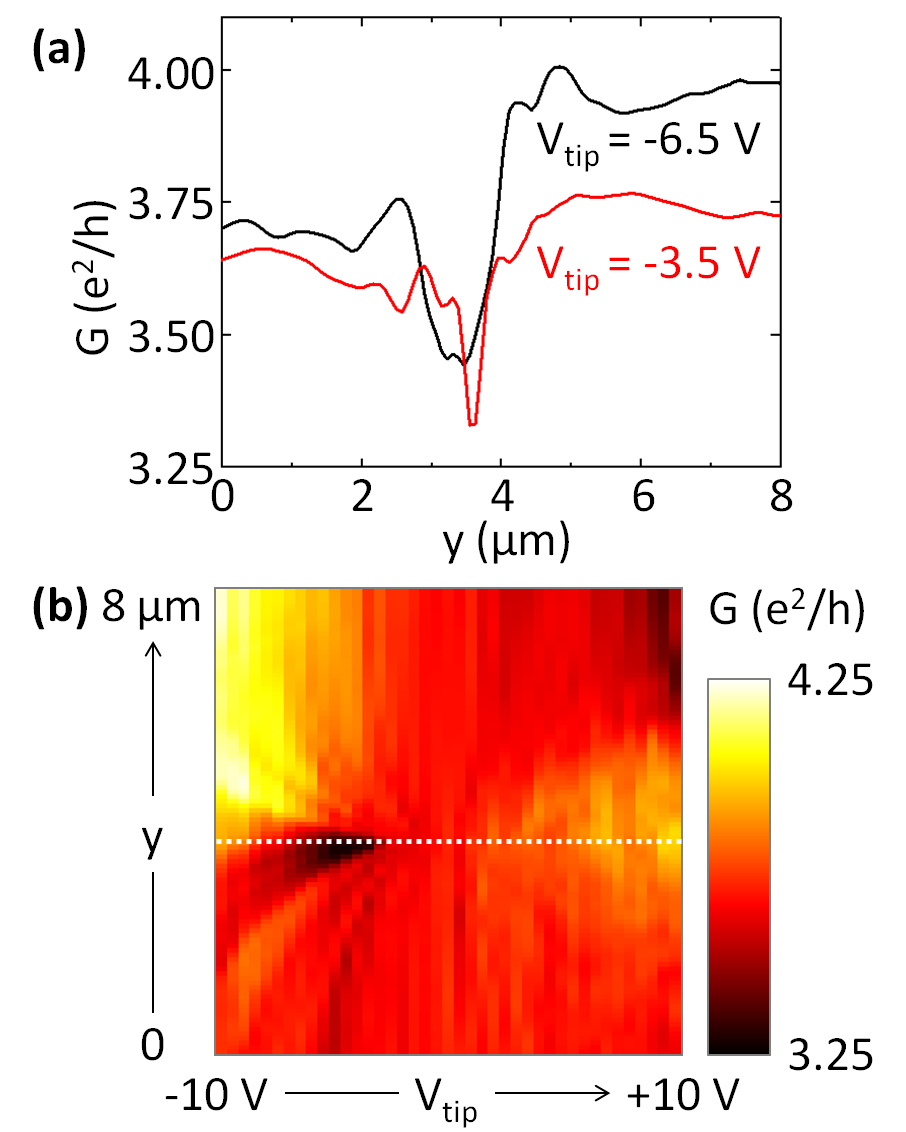}
		\caption{(Color) \textbf{(a)} Line scans (line indicated in Fig.~3a) for $V_{tip}=-6.5$~V (black upper trace) and $V_{tip}=-3.5$~V (red lower trace) show strongest conductance modulation $\Delta G \approx 0.5 e^2/h$ and most localized gating effect $\Delta y \approx 0.125~\mu$m (HWHM), respectively. \textbf{(b)} Evolution of line scans as a function of $V_{tip}$.}
	%\label{Fig4}
\end{figure}

Fig.~4b shows the conductance measured along the line marked in Fig.~3a as a function of $V_{tip}$. We observe resonant suppression of conductance corresponding to rings in the SGM maps mainly for negative tip voltages. As the tip voltage becomes more negative, the spatial separation between the conductance dips increases. This shift can be understood based on the electrostatics of the SGM configuration. When the tip voltage is changed, the potential at a particular sensitive site within the sample can be kept constant if the distance between tip and site is adjusted accordingly, as traced out by the bright and dark curves in Fig.~4b. More precisely, the tip voltage needed to trigger dephasing is more negative the farther away from the site the tip is, showing that the potential at the site needs to be pushed up for dephasing.
While the strong local gating effect with $|\Delta G| \approx 0.5 e^2/h$ can only be observed for $V_{tip}\leq-3.5$~V, weaker resonances with $|\Delta G| < 0.1 e^2/h$ can still be observed for smaller negative tip voltages. In the slightly $p$-conducting regime at $V_{back}=0$, backscattering at the three identified scattering sites occurs for tip voltages as small as $V_{tip}\approx-1.0$~V (sites 1 and 2) or $V_{tip}\approx-2.5$~V (site 3), whereas detectable backscattering cannot be induced in the adjacent regions for tip voltages as strong as -10~V. This shows that the potential at the sensitive sites need to be pushed up less to trigger dephasing than would be needed in surrounding regions.

While there might be quantitative differences between backscattering in the QSH regime ($V_{back}=+200$~V) and in the presence of weak bulk transport ($V_{back}=0$), respectively, we will use $V_{tip}\approx -12.5$~V, which was necessary in the QSH regime to introduce backscattering in otherwise unperturbed regions of the edge, as an estimate for the strength of a potential perturbation required to cause backscattering. This corresponds to a locally induced density $p\approx 2.5\times10^{11}~\rm{cm}^{-2}$~\cite{Supp}. Let us assume that densities associated with backscattering at the identified sites are also around $p=2.5\times10^{11}~\rm{cm}^{-2}$.  But the low tip voltages necessary to turn on backscattering at those sites correspond to tip-induced density changes only on the order of several $10^{10}~\rm{cm}^{-2}$. This implies that the metallic puddles at the scattering sites are already present in the absence of the tip for $V_{back}=0$. In addition, tuning the sample into the QSH regime by applying $V_{back}=+200$~V changes the bulk density by $|\Delta p| \approx 2\times10^{10}~\rm{cm}^{-2}$, i.e., significantly less than the inferred carrier density in the dephasing regions, indicating that the resonant backscattering observed in the QSH regime at $V_{back}=+200$~V can be attributed to $p$-type puddles interrupting the edge state.

All observed scattering sites can be identified as $p$-type whereas experimental signatures of pre-existing $n$-type puddles are absent. This may be related to the band structure of the HgTe structures: hole densities of a few $10^{11}~\rm{cm}^{-2}$ can already occur in the quantum well if the Fermi level is just 1~meV below the valence band maximum, much less than the shift into the conduction band required to induce comparable $n$-type densities~\cite{Supp}. This suggests that potential fluctuations of a few~meV might be responsible for the observed metallic scattering sites.
As our measurements imply that densities on the order of $10^{11}~\rm{cm}^{-2}$ are required to cause substantial backscattering, a conductance $G\approx 0.5e^2/h$ carried by a helical edge state coexisting with a bulk carrier density $p\approx 2\times10^{10}~\rm{cm}^{-2}$ at $V_{back}=0$, as determined earlier, appears reasonable.

For positive tip voltages, the pattern of conductance modulation in the SGM maps gets significantly more complex (Fig.~3d~-~f) so that attribution of individual features to particular scattering sites is rarely possible. The clearest features at the mesa edge are centered around site 3, where circular patterns are visible for both positive and negative tip voltages, suggesting that the site is not simply the location of a pre-existing potential fluctuation. 

A variation of the well width by just a single monolayer can change the bulk 2D energy gap by several meV, significant compared to the 13~meV gap predicted for our 8~nm quantum well~\cite{Supp}. This could naturally lead to a small region where local gating could enable both $p$- and $n$-type puddles at the same location. At the same time, a variation of the bulk gap could also result in a relative shift of the bands, locally shifting the Fermi level from the gap into one of the bulk bands.

\section*{Nature of bulk transport at low densities}

When we position the tip over the slightly $p$-conducting bulk at $V_{back}=0$, we see an overall trend from high conductance $G$ for negative tip voltages to low conductance for positive tip voltages (Fig.~3\textbf{a-f}). The increase in $G$ for negative tip voltages (Fig.~3\textbf{a-c}) can be understood as due to a tip-induced accumulation of $p$-type bulk carriers. For positive tip voltages (Fig.~3\textbf{d-f}), the conductance is suppressed by approximately $0.5 e^2/h$ compared to the conductance in the absence of a tip effect. Though the sign of this effect makes sense, its magnitude is several times stronger than one would expect for full suppression of bulk transport through only a small fraction of the device - probably around a few percent of the device width - in the case of a homogeneously conducting bulk with $G_{bulk}<4e^2/h$. But this assumption of homogeneous conduction is not justified.
If we assume a conductance $G\geq0.5 e^2/h$ for each edge state in the slightly $p$-conducting device at $V_{back}=0$ as determined above, we obtain a bulk resistivity exceeding a few hundred~k$\Omega/\Box$, a range characteristic of percolative or hopping bulk transport. In this transport regime, signatures of localized states are visible in the SGM maps not just along the edge of the device, but also in the bulk several microns away from the edge (Fig.~3g~-~l). In contrast to the superficially-similar sets of rings centered at the edge, however, the localized states in the bulk manifest themselves in a variety of conductance patterns. For negative tip voltages (Fig.~3g~-~i), for example, a set of concentric sharp rings with \textit{enhanced} conductance is visible in the upper right corner of the scan window, marking resonantly enhanced transport, as is typical for a quantum dot~\cite{Pioda04, Bleszynski08, Schnez10, Huefner11}. In the scans with $V_{tip}\geq+6.5$~V, a single elongated region of enhanced conductance emerges at the upper edge of the scan window and grows with increasing tip voltage (Fig.~3j~-~l). Such a response to a locally induced potential is reminiscent of the behavior seen in SGM studies of quantum point contacts~\cite{Schnez11}. The enhancement of conductance with increasingly positive tip voltages implies that the tip induces $n$-type carriers. This is very plausible, as $p$-type densities extracted above for negative tip voltages are significantly larger than the $p$-type background density in the absence of the tip, and the small band gap of HgTe quantum wells allows for a tip-induced shift of the Fermi level from the valence band into the conduction band. For large positive tip voltages, a set of sharp arcs of suppressed conductance (Fig. 3l) could point to the presence of another localized state strongly affecting transport in our device. The center of these arcs appears to be in the bulk outside the scan window, but probably still inside the channel in which we measure conductance. All these observations are consistent with a two-dimensional system with strong potential disorder, in which quantum dots and quantum point contacts can form accidentally as the Fermi level fluctuates between bands and gap. 

\section*{Concluding remarks}

The key results of our experiments are evidence of edge state transport in the QSH regime, identification of pre-existing scattering sites and demonstration that backscattering of the QSH states caused by these sites can be enhanced resonantly if the local perturbation is tuned appropriately. The separation between these well-localized scattering sites is typically between 1.5 and 2 microns, and the edge states appear to propagate unperturbed between them. This spacing qualitatively explains the size limit of 1 to a few microns for ballistic transport in QSH devices determined in earlier transport experiments~\cite{Koenig07, Koenig08, Roth09}, while also suggesting that if scattering sites could be individually tuned off-resonance the ballistic length could be extended substantially. The detection of localized states in the bulk supports our interpretation that the scattering sites along the edge can be attributed to local disorder which is inherent to the quantum well structure and not a result of damage imposed on the edge of the device during the fabrication process. In addition, we have demonstrated that backscattering can be introduced in otherwise unperturbed QSH states by a sufficiently strong tip-induced perturbation. This gives insight into the robustness (or vulnerability) of the QSH states against potential fluctuations. However, as the conductance modulation solely induced by the tip potential is weaker than the effect of the pre-existing scattering sites identified in our current experiments, it would be worthwhile to study these effects in more detail in subsequent experiments. Similar SGM experiments in an external magnetic field could elucidate the role of time-reversal symmetry for the predicted suppression of backscattering.\\

\begin{acknowledgements}
We thank M.R.~Calvo, J.~Maciejko, K.~Moler, K.~Nowack, and T.~Schmidt for valuable discussions. This work was funded by the Department of Energy, Office of Basic Energy Sciences, Division of Materials Sciences and Engineering, under contract DE-AC02-76SF00515 (SGM and transport investigation of the QSH state in HgTe devices), by the Center for Probing the Nanoscale, an NSF NSEC, supported under Grant No. PHY-0830228 (development of the SGM technique), and by the DARPA Meso project under grant no. N66001-11-1-4105 (MBE growth of the HgTe heterostructures). M.K. acknowledges financial support from the DAAD Postdoctoral Fellowship (Deutscher Akademischer Austausch-Dienst, German Academic Exchange Service).
\end{acknowledgements}

\setcounter{figure}{0}
\renewcommand{\thefigure}{S\arabic{figure}} 
\section {Supplemental Material}
\section*{Device characterization}
The two-terminal configuration used in the SGM experiments is not well-suited for determining the sample properties like carrier density and mobility. For that purpose, we fabricated a Hall bar structure with $L \times W = 50~\mu\rm{m}\times30~\mu\rm{m}$ from the same material. Like in the SGM device, we used a metallic layer at the bottom of the CdTe substrates (thickness $\sim 800~\mu$m) in our samples as a back gate electrode. When a voltage is applied to this back gate electrode to tune the Fermi level in the device, the Hall bar enters the QSH regime around $V_{back} = 0$ and $n$-type bulk conduction occurs for $V_{back}>100$~V (Fig.~S1a). The conductance value of $G \approx 0.15 e^2/h$ for the QSH state is comparable to values reported for other devices of similar size \cite{Koenig07, Koenig08}. In the vicinity of the conductance minimum, several sharp peaks can be observed. These might be caused by resonances in the scattering mechanism as they have been observed in our SGM data, now with the back gate electrode tuning the properties of multiple dephasing regions simultaneously.

The voltage range for which the Hall bar is tuned into the QSH regime is shifted by $\Delta V_{back} = -150$~V in comparison to the SGM device. Based on a parallel plate capacitor model, this corresponds to a difference in density of approximately $1.5\times10^{10}~\rm{cm}^{-2}$ toward lower p-type densities in the Hall bar. Such shifts in density have been observed in several devices. They can be explained by fluctuations of the carrier density on length scales much larger than the typical device size and, to a lesser extent, by small variations in the sample properties between different cooldowns.

\begin{figure}[htb]
	\includegraphics[width=\linewidth]{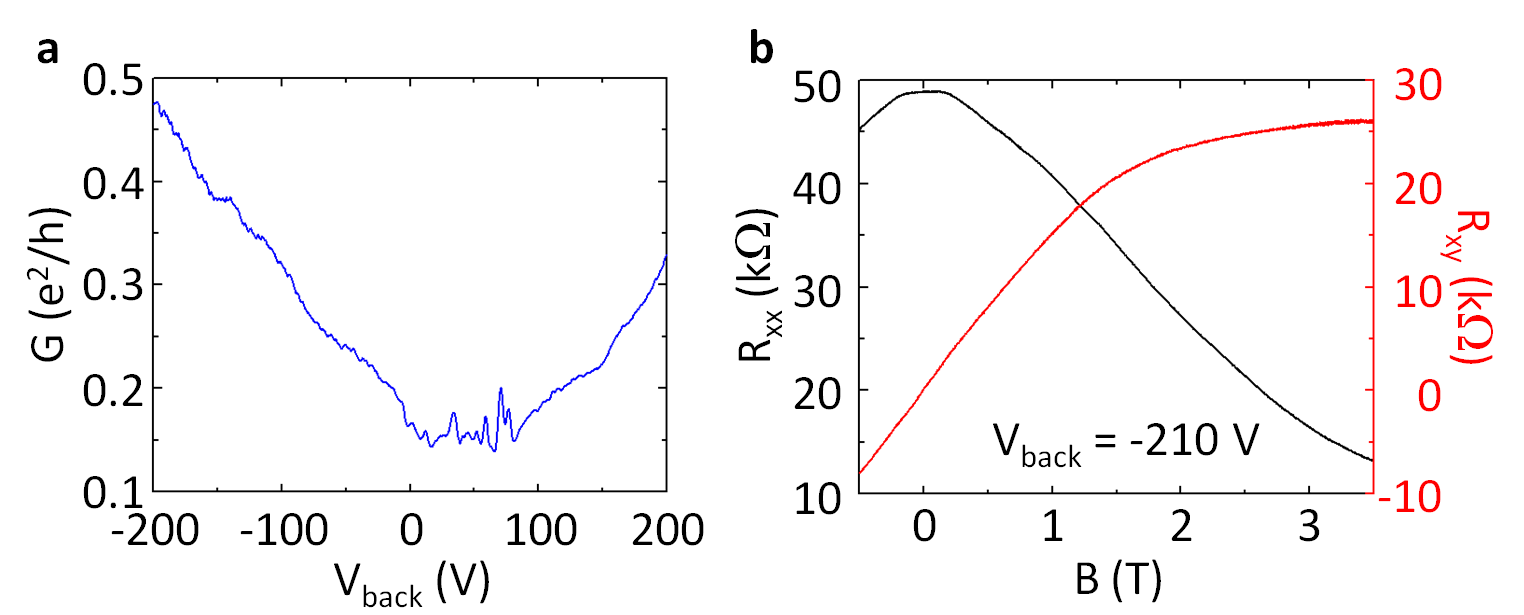}
	\caption{\textbf{(a)} The Fermi level in a Hall bar device can be tuned from the valence band through the gap into the conduction band. \textbf{(b)} For a large negative back gate voltage $V_{back}=-210$~V, the device is clearly p-type bulk conducting and carrier density and mobility can be determined from magnetotransport measurements.}
	%\label{FigS1}
\end{figure}

Magnetotransport measurements were performed to determine the density and mobility in the quantum well (Fig.~S1b). In close proximity to the QSH state, the edge states contribute significantly to the transport, making extraction of bulk properties from magnetotransport data difficult. For $V_{back}=-210$~V, the sample has a $p$-type carrier density and mobility of $p = 3.7\times10^{10}~\rm{cm}^{-2}$ and $\mu = 5400~\rm{cm}^2(Vs)^{-1}$, respectively. The clear reduction of carrier mobility in comparison to n-type quantum wells with comparable thickness \cite{Koenig07, Roth09, Bruene12} can be explained by the large effective mass in the valence band which is about one order of magnitude higher than the effective electron mass in the conduction band.

Based on these results, we estimate the carrier density in the SGM device to be $p\approx 2\times 10^{10}~\rm{cm}^{-2}$ at $V_{back}=0$. This estimate does not take into account uncertainties based on the large disorder in the sample, or on contributions of the edge state to the Hall resistance even in the presence of substantial bulk conduction at $V_{back}=-210$~V. The device geometry itself, in particular the short length of the device between the metallic contacts to the device, may also result in a gate response different than expected from the assumed parallel plate model. However, as the discussion in the main paper does not rely on the exact carrier density in the slightly $p$-type device, the given estimate is sufficient for our purposes.

\section*{Band structure calculations}

The typical analysis of subband energies in HgTe structures as a function of the quantum well thickness \cite{Koenig07, Pfeuffer} considers the energies of the subbands at $k=0$. For narrow HgTe quantum well layers, the band structure shows a direct gap at $k=0$ so that the given energies for the respective subbands represent the bulk gap correctly. However, already for $d_{QW}=8.0$~nm the band structure shows an indirect gap with a valence band maximum at $k\neq0$ \cite{Bruene12}. Thus, the separation of valence and conduction band at $k=0$ overestimates the bulk gap for wider quantum wells. We performed band structure calculations for HgTe quantum wells within a $8 \times 8~\textbf{k}\cdot \textbf{p}$ model \cite{Novik05} to determine the exact gap size in HgTe QW structures with a layer sequence identical to our samples (Fig.~S2a). For QW layers with a thickness up to $d_{QW}\leq 7.0$~nm, the band structure has a direct gap at $k=0$. For wider QW layers, the inverted gap becomes indirect and reaches its maximum size of $E_{gap} = -13$~meV at $d_{QW}=8.0$~nm (Fig.~S2b). For even wider quantum wells, the gap size decreases continuously until a transition to a semimetallic band structure occurs around $d_{QW}=15$~nm. The semimetallic behavior was recently observed experimentally in 20~nm wide quantum wells \cite{Olshanetsky12}.

\begin{figure}[htb]
	\includegraphics[width=\linewidth]{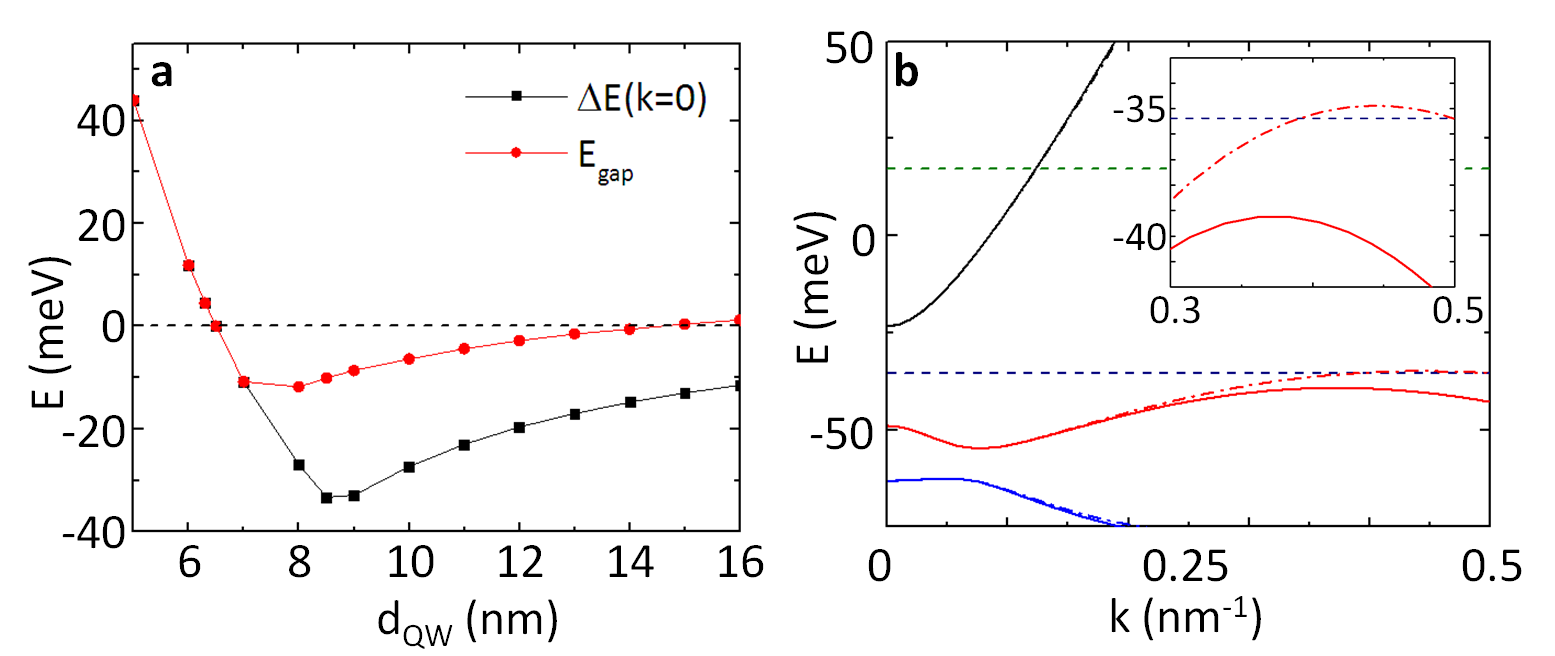}
	\caption{\textbf{(a)} The evolution of the bulk gap in HgTe quantum wells as a function of the QW width was determined by $8 \times 8~\textbf{k}\cdot \textbf{p}$ band structure calculations. \textbf{(b)} Dispersion for a HgTe structure with $d_{QW}=8$~nm as it was used in our SGM experiments. The dashed horizontal lines indicate the Fermi energy for $n=2.5\times10^{11}~\rm{cm}^{-2}$ (green) and $p=2.5\times10^{11}~\rm{cm}^{-2}$ (blue), respectively. Inset: Valence band maximum and Fermi level for $p=2.5\times10^{11}~\rm{cm}^{-2}$.}
	%\label{FigS2}
\end{figure}

The flat dispersion of the valence band near its maximum at finite $k$ allows for large carrier densities already for a Fermi level only a few meV below the valence band maximum whereas much higher Fermi energies, now with respect to the minimum of the conduction band, are required to obtain comparable $n$-type densities (Fig.~S2b).

\section*{Full suppression of edge conductance}

In the main text, we showed that the conductance of the edge state can be fully suppressed by local gating (see Fig.~2). However, as we have seen such a strong effect of the tip potential on the conductance only in one case, our data is not sufficient to extract details regarding the underlying mechanism. In the following, we will discuss several proposed models that in principle can result in a full suppression of transport in a QSH edge state and assess their applicability in our samples. 

The first possible scenario relies on coupling between the QSH edge and a nearby magnetic impurity \cite{Maciejko09}. While conventional magnetic impurities are not expected to be present in our samples, an accidentally-formed quantum dot could play the same role. If the dot is occupied by an odd number of electrons, one electron spin remains unpaired and serves as a magnetic impurity. In ordinary 1D ``Luttinger liquids", coupling to a local site should fully suppress conductance as $T\rightarrow0$, even for weak electron-electron interaction. In contrast, in a QSH ``holographic liquid" at the 1D boundary of a 2D system, for weak electron-electron interaction a magnetic impurity should be screened by the formation of a Kondo singlet, restoring conductance of the QSH edge state at low temperatures. For strong electron-electron interactions, two-particle backscattering is predicted to result in the formation of a ``Luttinger liquid insulator" characterized by full suppression of edge state transport for $T\rightarrow0$. The strength of electron-electron interaction can be defined by the Luttinger parameter $K$ for the helical liquid \cite{Maciejko09, Kane92} with $0 \leq K \leq 1$ where lower $K$ connotes stronger interaction. While it is difficult to determine $K$ for a given system exactly, it can be estimated when the relevant interactions are taken into account \cite{Kane92}. In our specific case, the strength of Coulomb interaction in the HgTe devices depends on the device geometry: $K\approx0.55$ for the devices with a top-gate electrode as used in earlier transport experiments \cite{Koenig07, Koenig08, Roth09, Bruene12}, whereas the absence of such an electrode in our SGM devices leads to a reduced screening of electron-electron interaction and the Luttinger parameter decreases to $K\approx0.35$. This value is close to the regime of $K<0.25$ required for the ``Luttinger liquid insulator" and a small error - either inherent to the approximation or due to a slight deviation of the sample parameters used in calculating $K$ - might result in an overestimation of $K$ in our devices. Thus, interactions in our devices might be stronger than anticipated. However, it is not clear whether our experimental temperature should be low enough to see nearly full suppression under this scenario.

Rashba spin-orbit interaction has also been proposed to lead to localization of QSH edge states if it is spatially nonuniform along the edge \cite{Stroem10}. HgTe quantum well structures show very strong spin-orbit interaction, and a Rashba splitting in the conduction band of up to 30~meV \cite{Zhang01,Gui04, Hinz06}. In addition, the Rashba interaction can be tuned externally by an electric field, making the proposed model a seemingly good candidate for an explanation of the observed behavior. However, the proposed Rashba-induced localization length shows a strong dependence on the electron-electron interaction strength \cite{Stroem10}. For $K\approx0.35$, the localization length would exceed the size of our device, so the full suppression we observe cannot be explained by spatially inhomogeneous Rashba interaction in this case. To induce localization on a scale of order 100~nm, as we infer it to happen in our sample, $K<0.2$ is required which is significantly smaller than the estimated value $K\approx0.35$ based on our device properties.

Although we apply no magnetic field in our measurements and use no magnetic materials in the fabrication of the devices, for completeness we note that backscattering becomes possible when time-reversal symmetry is broken. Several theoretical proposals have considered the effect of a magnetic field, applied either locally or globally, with a resulting gap in the linear Dirac spectrum of the QSH edge states. Two similar models \cite{Qi08, Timm11} considered magnetic field applied locally along the edge to form a quantum dot in the QSH edge channel, suppressing conductance when the dot is tuned into Coulomb blockade. More recently, a local potential perturbation was predicted to suppress transport in the QSH state in the presence of an external magnetic field \cite{Ilan12}. However, with no magnetic field in our experiments, these mechanisms cannot explain the observed full suppression either.

\begin{figure} [hbt]
		\includegraphics[width=\linewidth]{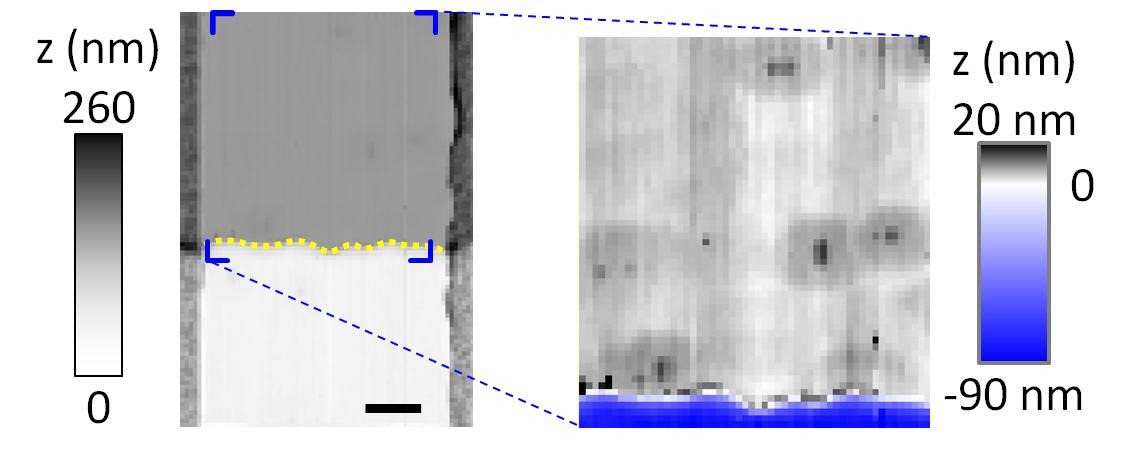}
		\caption{Defects on the mesa surface are clearly visible in a high-resolution in-situ AFM scan. The scale bar in the AFM on the left hand side (same as Fig.~1(a) in the main paper) is 1 micron.}
	%\label{FigS3}
\end{figure}

While existing theoretical models do not appear to account for the full suppression of edge state transport, the observed behavior might be related to the presence of strong defects in our samples. As can be seen in Fig.~S3, multiple defects of a characteristic shape can be found within an area of tens of $\mu\rm{m}^2$ on the mesa surface. These defects typically have a circular shape with a diameter of approximately 1 micron and a height of a few nanometers. Several of these defects feature a central ridge, occasionally elongated along one of the $<110>$ directions, with an additional height of approximately 10~nm. One of these particularly tall defects can be found at the center of the region of strong suppression shown in Fig.~2 in the main paper. We believe that these defects originate at the surface of the substrate and subsequently propagate through the MBE-grown layers of the heterostructure. The resulting lattice distortion can affect the layer structure in multiple ways. If the quantum well thickness locally is reduced below the critical value $d_{QW, crit}\approx6.3$~nm, the affected region would constitute a topologically-trivial insulator. A trivially insulating region could also occur if the quantum well is completely interrupted by the defect (which is taller than $d_{QW}=8$~nm) or if the well and barrier material, respectively, get intermixed \cite{Li09} due to imperfect growth conditions in the vicinity of the defect. In either case, the QSH edge state should propagate along the boundary between topologically trivial and non-trivial insulator. The creation of trivially-insulating regions may also cause a strong deformation of the edge states which could form loops in the edge state \cite{Vaeyrynen11} or antidots of helical edge states coupled to the QSH state at the edge of the mesa. Such perturbations of the helical edge state likely will cause some degree of backscattering, but it is expected to be a weak effect, not full suppression \cite{Vaeyrynen11}. On the other hand, a strong local increase in the QW thickness could result in a semimetallic band structure (see above and Ref.~\citenum{Olshanetsky12}), resulting in a bulk-conducting region coupling to the QSH state. Finally, the defect could also cause a local shift of the chemical potential relative to the band edge, possibly into the bulk bands. In either case, the resulting metallic puddle could couple to the edge state and cause a suppression by up to $0.5e^2/h$, but again no full suppression \cite{Roth09}.

In summary, it is not clear that the models already discussed theoretically suffice to explain our observed full suppression (and some can even be ruled out completely as explanations in our case). Nor do the obvious consequences of a strong defect appear to give rise to a full suppression. Thus, it will require further effort to understand the observed full suppression caused by local gating near the edge.

\section*{Simulating the tip-induced potential profile}

We used COMSOL\cite{Comsol} to calculate the electrostatic potential profile of the tip and the resulting induced modulation of the carrier density. We simulated the effect of a conical tip over a quantum well structure with a lateral extension exceeding the expected long-range effect of the tip. An axially symmetric layout was chosen to reduce the computational complexity. This simplified configuration will not provide quantitatively exact results as it does not take into account details of our device geometry such as the edge of the heterostructure, giving rise to a spatially varying dielectric environment, and the nearby Ohmic contacts which might shield the tip potential. Nonetheless, the simulations can give a rather accurate representation of the tip-induced density profile. Fig.~S4a shows the density induced in the quantum well layer for $V_{tip}=-10$~V. At the center, i.e., directly beneath the tip, a hole density of $2.6\times10^{11}~\rm{cm}^{-2}$ is induced. The density profile decays rapidly with a full width at half maximum of 520~nm.

\begin{figure}[htb]
	\includegraphics[width=\linewidth]{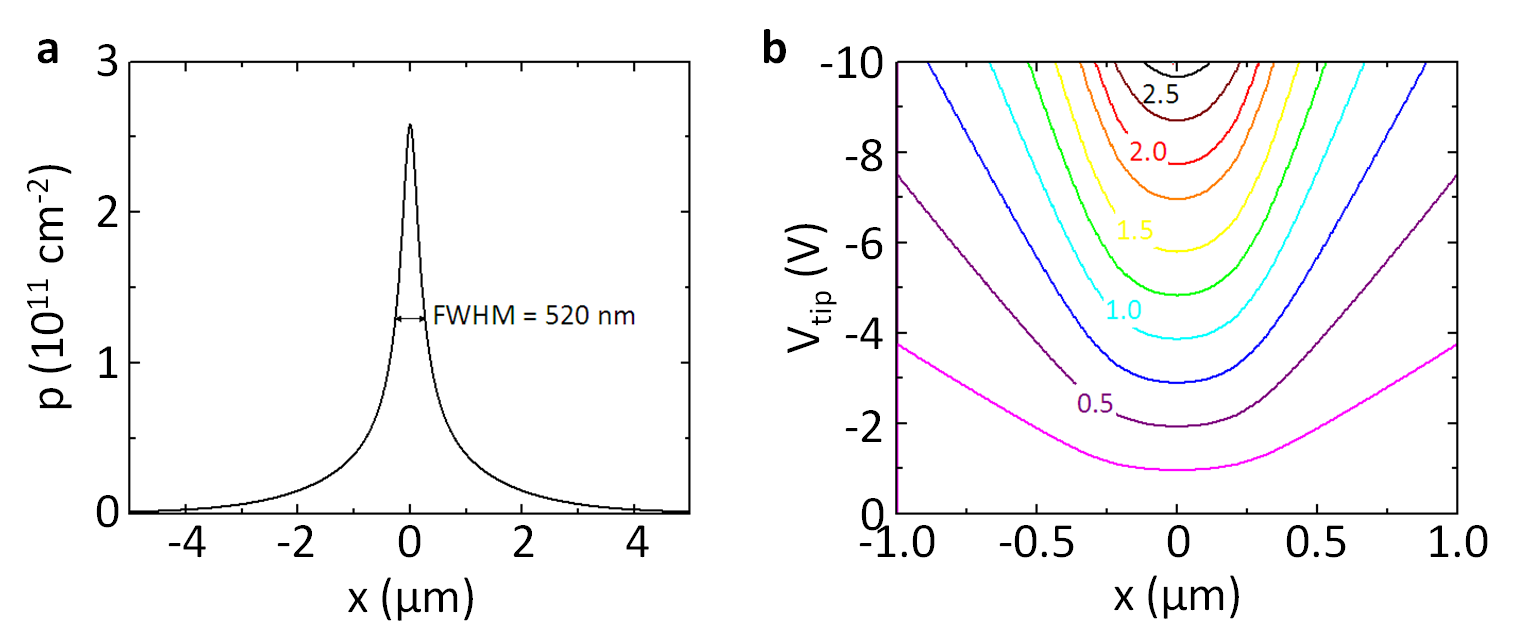}
	\caption{\textbf{(a)} Profile of the hole density induced for $V_{tip}=-10$~V. \textbf{(b)} Size of induced metallic regions with densities exceeding certain threshold densities (labels given in $10^{11}~\rm{cm}^{-2}$).}
	%\label{FigS4}
\end{figure}

If we assume that the carrier density in a finite-sized region translates directly into the strength of the dephasing mechanism, we can estimate how the size of a region with a given dephasing strength evolves as a function of the tip voltage. In Fig.~S4b, we plot the evolution of quasi-two-dimensional regions above a given threshold density as a function of the applied tip voltage. As discussed in the main paper, a density on the order of $10^{11}~\rm{cm}^{-2}$ is required to induce a measurable amount of backscattering. For the corresponding tip voltage of $V_{tip}<-10$~V, the diameter of the induced dephasing region is a few 100 nm which is comparable to our size estimate for the pre-existing scattering sites and the size of the metallic region studied theoretically in Ref.~\citenum{Roth09}. Thus, the simulations of the electrostatic potential profile further substantiate our interpretation of the SGM results. For simplicity, we neglect the likely effect that higher density at the center of the puddle will result in an increase of the dephasing strength.

\section*{Tip-induced emergence of bulk transport}

We already briefly discussed in the main paper and showed in Fig.~2a that conductance values exceeding $G=2e^2/h$ - the expected upper limit for the conductance in the QSH regime - can be observed when large negative tip voltages $V_{tip}\leq-12.5$~V are used to manipulate the edge states. We attribute this excess conductance to an emergence of bulk conductance caused by a long-range effect of the tip potential. Simulations show that densities on the order of several $10^9~\rm{cm}^{-2}$ can be induced even microns away from the actual tip position for such large tip voltages (Fig.~S5a). Our SGM devices are $5~\mu$m long so that they are affected along their entire length by the tip potential. Thus, the bulk region close to the tip can get tuned into the valence band, resulting in p-type bulk transport. Figs.~S5b and S5c show SGM maps taken with $V_{tip}=-14.5$~V and -13.5~V, respectively. It can be seen that a more negative tip voltage results in a larger bulk contribution to the total conductance as it is expected for tip-induced bulk conductance.

\begin{figure}[hbt]
	\includegraphics[width=\linewidth]{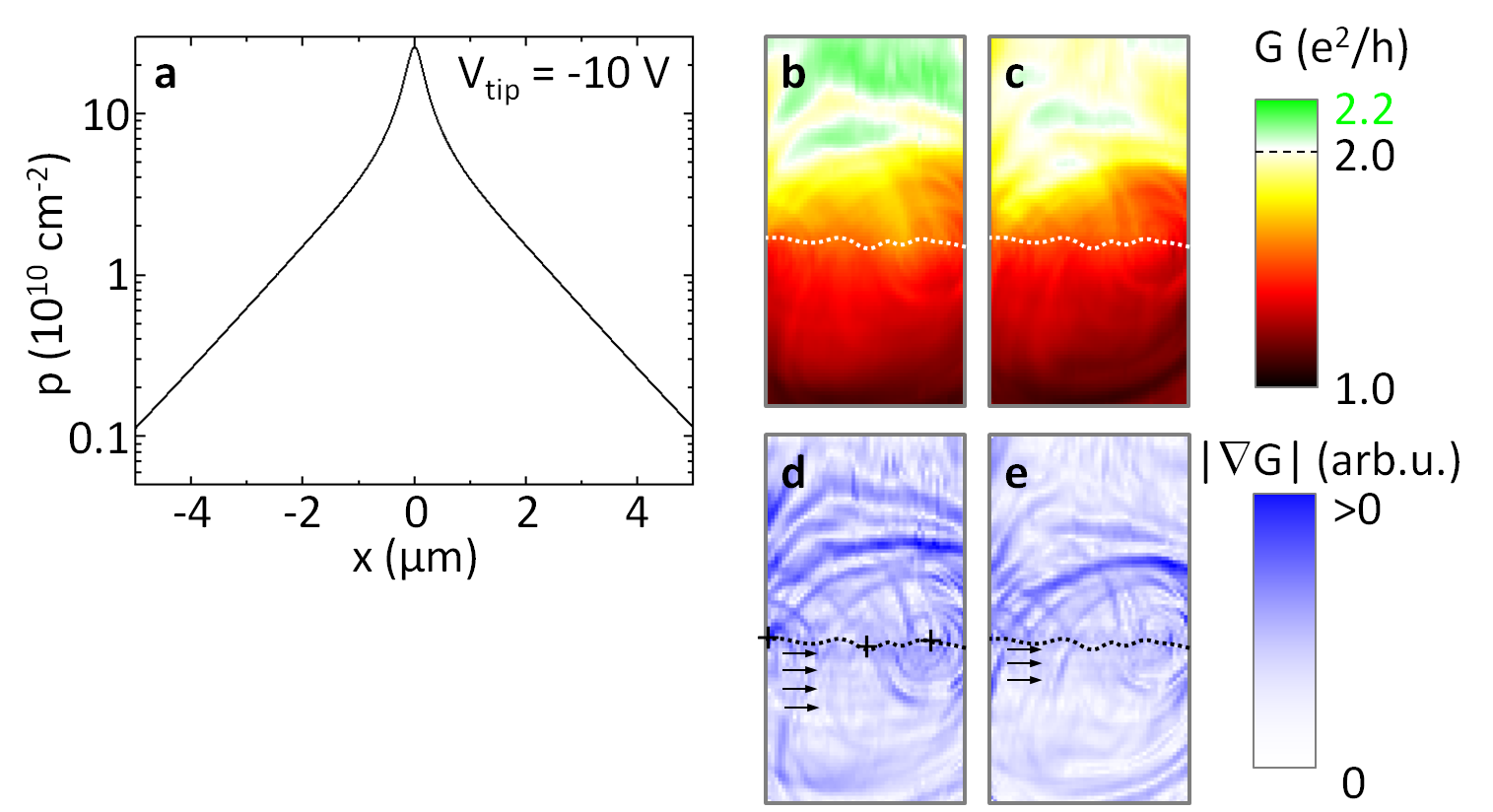}
	\caption{\textbf{(a)} Profile of the tip-induced density for $V_{tip}=-10.0$~V (same data as Fig.~S4a, now plotted log-scale). \textbf{(b), (c)} SGM conductance maps for $V_{tip}=-14.5$~V and -13.5~V, respectively. \textbf{(d), (e)} Gradient maps $|\nabla G(x,y)|$ of (\textbf{b}) and (\textbf{c}), respectively.}
	%\label{FigS5}
\end{figure}

In the maps plotting the gradient of the conductance (Figs.~S5d and e), parallel lines similar to the ones shown in Fig.~2e are visible. These additional maps clearly demonstrate the evolution of the lines as a function of the tip voltage: for more negative tip voltages, the lines are more widely spaced. This confirms that the solely tip-induced conductance modulation in the otherwise unperturbed sections of the mesa edge is a function of the potential at the position of the edge states which is consistent with the picture of dephasing occuring in finite-sized metallic regions. 

Future SGM experiments with coaxially shielded tips \cite{Harjee10} which produce a much more localized potential perturbation could help with many of these complications. In particular, we expect the long tails in the induced potential to be significantly suppressed relative to the potential directly below the tip. This improvement will consequently eliminate the long-range gating effect responsible for the emergence of bulk conductance and thus should allow for a more quantitative analysis of the observed conductance modulation.

%\bibliography{SGM}

\end{document}